\documentclass[%
 reprint,
 amsmath,amssymb,
 aps,
prl,
floatfix,
]{revtex4-2}

\usepackage{tikz-cd}
\usepackage{epstopdf}
\usepackage{graphicx}
\usepackage{dcolumn}
\usepackage{bm}
\usepackage{comment}
\usepackage{color}

\begin{document}

\preprint{APS/123-QED}

\title{Many-body delocalization from embedded thermal inclusion}

\author{J. Clayton Peacock}
\affiliation{Department of Physics, New York University, New York, NY-10003, USA}
\author{Dries Sels}%
\affiliation{Department of Physics, New York University, New York, NY-10003, USA}
\affiliation{Center for Computational Quantum Physics, Flatiron Institute, New York, NY-10010, USA}

\date{\today}

\begin{abstract}
We numerically study quantum avalanches in 1D disordered spin systems by attaching two XXZ spin chains. One chain has low disorder representing a rare Griffith's region, or thermal inclusion, and the second has larger disorder, i.e. disorder larger than the observed finite-size crossover. Comparing dynamics of this system to identical systems with uniformly large disorder, we find evidence for exponentially slow thermalization (in disorder) within the MBL regime when the rare region is present. We observe a decay of the spin imbalance in the bulk of the large disorder region that persists to long times ($\sim10^{4}$) and find a universal behavior of the spectral function. 
\end{abstract}

\maketitle


\section{\label{sec:level1}Introduction}
The existence of a many body localized (MBL) phase in interacting, disordered spin chains was first considered by Anderson~\cite{Anderson}. While he focused on the dilute limit, arguing that this could be treated as a non-interacting problem, it remained unclear for many decades if the more generic, interacting version of this phase existed. 
Intensive research on this MBL phase has progressed, particularly after evidence of its possible stability at high temperatures was reported in the early 2000s \cite{BASKO20061126,gornyi2005MBL,Oganesyan2007}. Given local interactions of strength $J$ and lattice disorder strength $W$, an MBL system is thought to undergo a phase transition from a thermal to an integrable regime at a large enough critical disorder $W_c$ on the order of a few $J$. Above this critical disorder, MBL would be an emergent integrable phase of matter violating the eigenstate thermalization hypothesis (ETH) \cite{Srednicki1994ETH,Dalessio2015FromQC}, characterized by the existence of an extensive set of local integrals of motion (LIOMs), or ``L-bits'', which are related to single-site spin operators by a unitary rotation \cite{annurev-conmatphys-031214-014726,Huse2014Phenomenoloy,Imbrie2016,ROS2015420,Serbyn2013,Thomson2018}. This insulating phase has been studied extensively in previous works \cite{AbaninReview,annurev-conmatphys-031214-014726,ALET2018498}, and although there is experimental \cite{Schreiber2015,Rubio-Abadal2019,Smith2016,Luschen2017,Kohlert2019,Lukin2019,Rispoli2019} and numerical evidence of the existence of a nearly localized regime in smaller system sizes at large enough disorder (estimates of which have varied around $W_c\approx3-6$) \cite{Pal2010MBLTransision,Lev2015,Luitz2015,Wahl2017,Mierzejewski2016,Mace2019,Chanda2020,Laflorencie2020,Gray2018,Doggen2018,POLFED2020,Serbyn2015MBLTransition,Sierant2020ThoulessTime}, there is also increasing evidence in recent years that the MBL phase has to be unstable in the thermodynamic limit in this range of disorder\cite{Leonard2023,suntajs2019quantum,suntajs2020transition,Kiefer-Emmanouilidis2021,Dynamicalobstruction,Bath-induceddelocalization,Sels2021}. \par

Exact numerical studies of interacting many body systems are limited to small system sizes $L$. Tensor network methods such as time-evolving block decimation (TEBD) \cite{Vidal2003,Vidal2004} or the time-dependent variational principle (TDVP) \cite{TDVP,Haegeman2016,Goto2019,Koffel2012} allow for time evolving larger systems, giving access to larger sizes of the order $L\approx200$ \cite{Challenges,Chanda2020,Doggen2018} but are often limited in the accessible simulation times.
Crucially, one of the destabilizing forces of MBL in the thermodynamic limit, so called ``quantum avalanches''~\cite{DeRoeck2016_AbsenceofMobilityEdges,deRoeck17,DeRoeck2017_StabilityandInstability_Imbrie,Thiery2018_MBDelocalizationasQuantumAvalanche,Gopalakrishnan2019_InstabilityofMBLasPhaseTransition,Goihl2019}, cannot be captured by small system size simulations in typical MBL spin chains. A quantum avalanche originates in the rapid growth of entanglement inside rare Griffith's regions \cite{Agarwal2016,Griffiths1969,Vojta2010} of very low disorder within the chain, which are sure to exist in the thermodynamic limit for purely statistical reasons, no matter how large the chain's average disorder. The probability of such a rare region is exponentially small in its length $\ell$, and one would expect a minimal $\ell\sim W/J$ to be required to make sure that there are some many-body states to resonate with. Therefore these regions do not occur in the sizes accessible by current numerical techniques. \par
As these regions are sure to exist at arbitrary sizes and potential strengths in the thermodynamic limit, the question is whether they thermalize the surrounding spins at a fast enough rate to take over the entire chain and thermalize the system. Many previous studies have investigated the avalanche mechanism using a variety of idealized analytical and numerical models in which rare regions are connected to perfectly localized LIOMs \cite{Chandran2016_MBLbeyondEigenstates,deRoeck17,Luitz2017_HowaSmallQuantumBathCanThermalize,Ponte2017_HowOneSpin,Crowley2020_AvalancheInducedLocalizedandThermalRegions,Potirniche2019_ExplorationofStabilityofMBL,Crowley2022_MeanFieldTheoryofFailedThermalizingAvalanches,Suntajs2022_ErgodicityBreakinginZeroDimensions}. These studies substantiate the idea of an avalanche-induced ergodicity-breaking transition when LIOMs are exponentially localized and give evidence that avalanches will induce thermalization at any disorder for spatial dimensions larger than one. To study the stability of MBL in a truly one-dimensional model, subsequent studies connected a 1D disordered spin chain to a Markovian bath \cite{MorningstarAvalanches, Bath-induceddelocalization}, representing a rare region, and found evidence of thermalization up to $W=20$ \cite{Bath-induceddelocalization}.
A natural next step is to investigate if an apparently localized strong disorder MBL region is also destabilized by the presence of a  more realistic, finite chain of lower disorder, representing a rare region as it would appear in the thermodynamic limit of a one-dimensional spin chain (see also Ref.~\cite{sarang15} and Ref.~\cite{Goihl2019}). As one cannot expect to see such a rare region arise naturally in accessible systems, such a region must be introduced by hand. In this work, we report on the stability of the MBL phase at large disorder in the presence of a ``rare region'' spin chain of much lower average disorder.

\section{\label{sec:level2}Embedded Inclusion Model}
Consider the 1D disordered Heisenberg model:
\begin{equation}\label{XXZ_H}
H=\sum_{i=1}^{N-1} \left(S^X_{i}S^X_{i+1}+S^Y_{i}S^Y_{i+1}+S^Z_{i}S^Z_{i+1} \right)+ W\sum_{i=1}^{N}h_iS^z_i 
\end{equation}
Here $h_i$ are i.i.d. uniformly distributed random numbers in the range $[-1,1]$ and $W$ is the disorder strength of the on-site potentials. In contrast to typical models of disordered spin chains, our system consists of two smaller chains $A$ and $B$, each with different disorder strength $W_{A,B}$, as shown in Fig.~\ref{fig:diagram}. Chain $A$, with length $L_{A}=12$, is designed to behave like a typical large disorder MBL chain. Chain $B$, with length $L_{B}=10$, is designed to behave like a rare region having only weak disorder $W_{B} = 1/2$. 
The size of the rare region $L_{B}$ is chosen to be sufficiently large such that even at the largest disorder studied here, i.e. $W_{A}=12$, the bath has a large enough bandwidth to resonate with boundary spins. If $L_B$ is too small, then there is no observable difference with unbiased sampling (as shown in Ref. \cite{Goihl2019}). 
As a reference point we also consider the model without the rare region, with $W_{B} = W_{A}$, which we refer to as the unbiased case. We will probe this model in two different ways: (i) by studying the imbalance in a kicked version of the problem where the chains are connected and disconnected periodically and (ii) by studying the spectral function in a brickwork quantum circuit approximation of the problem.  

\section{Block evolution}
\label{sec:Block}
\begin{figure}[t]
\includegraphics[width= 0.45\textwidth]{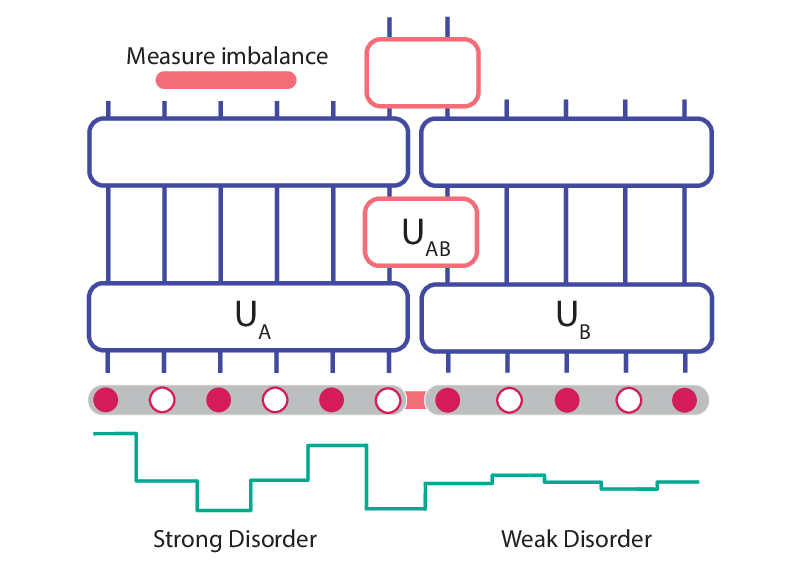}
\caption{\label{fig:diagram} Diagram depicting the general setup, where we connect a region $A$ with strong disorder to a region $B$ with much weaker disorder. The system is initialized in a staggered spin configuration, after which we evolve the system in a circuit fashion, by applying unitaries $U_A$ and $U_B$ to the separate chains alternated with a Heisenberg coupling gate $U_{AB}$ that couples the chains. We measure the imbalance in the strong disorder region.}
\end{figure}

To time evolve this system we implement a TEBD-type algorithm which operates by evolving each of the two chains separately and alternated with a connection step, see Fig.~\ref{fig:diagram}. Specifically, we use unitaries $U_{A,B}=e^{-i\tau H}$, with the Hamiltonian $H$ given by expression~\ref{XXZ_H} and $\tau=10$. The connection unitary $U_{AB}$ is a single Heisenberg gate connecting the ends of both chains. It should be noted that the time-step $\tau$ is rather large, meaning that information can spread throughout the rare region in between two consecutive connection steps. While this seems to somewhat speedup the thermalization process, similar results will be obtained in the brickwork setup described in the next section. The main reason for using large enough timesteps is purely numerical, i.e. it allows us to access a larger total timewindow for the same computational time. This is crucial, as the thermalization (if there is any) is expected to be incredibly slow. 

As a measure of thermalization, we consider the standard spin imbalance as a function of simulation time, $I(t)$, calculated over various sites $3-9$ in the typical MBL region: 
\begin{equation}\label{Imb}
I(t)=D\sum_{i=3}^{9}\langle\psi(t)|S^z_i|\psi(t)\rangle\langle\psi(0)|S^z_i|\psi(0)\rangle
\end{equation}
And, we initialize the XXZ chain in the far-from-equilibrium and zero magnetization N\'eel state $\left| \psi(0)\right>=\left|\uparrow,\downarrow,\uparrow,\cdots,\downarrow \right>$. Note that we start at the third site and end at the ninth to avoid boundary effects, and $D$ is chosen such that $I(t=0) = 1$. This spin imbalance, being a combination of local observables, is a measure of the retained local information about a system's initial state. In ergodic systems with eigenstates obeying the ETH, one expects the imbalance to decay diffusively to some $O(1/N)$ value as local information is transferred to nonlocal degrees of freedom \cite{annurev-conmatphys-031214-014726}. For an integrable or fully localized system, one instead expects the imbalance to converge to a constant value at long times, indicating the existence of a set of conserved quantities preserving local information about its initial state. The behavior of the imbalance in the non-interacting (Anderson) case and traditional uniform disorder XXZ chains of many sizes has been extensively reported (see for example, Ref.~\cite{Challenges,Luitz2016ExtendedRegime}), and thus this work will focus on its behavior in the presence of a thermal inclusion. Results in this section are obtained by evolving matrix product states (MPS) using the Julia ITensor library \cite{itensor} and the convergence is discussed in appendix A. 

\begin{figure}[t]
\includegraphics[trim=5 0 25 40,clip,width=0.48\textwidth]{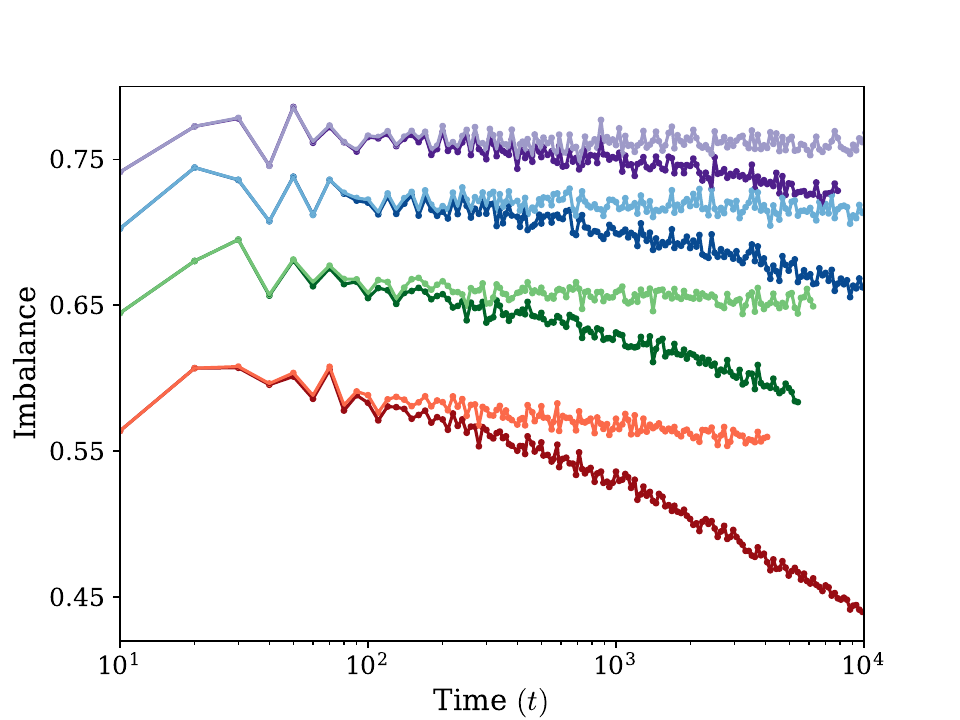}\par
\vspace{.5cm}
\includegraphics[trim=5  0 25 40,clip,width=0.48\textwidth]{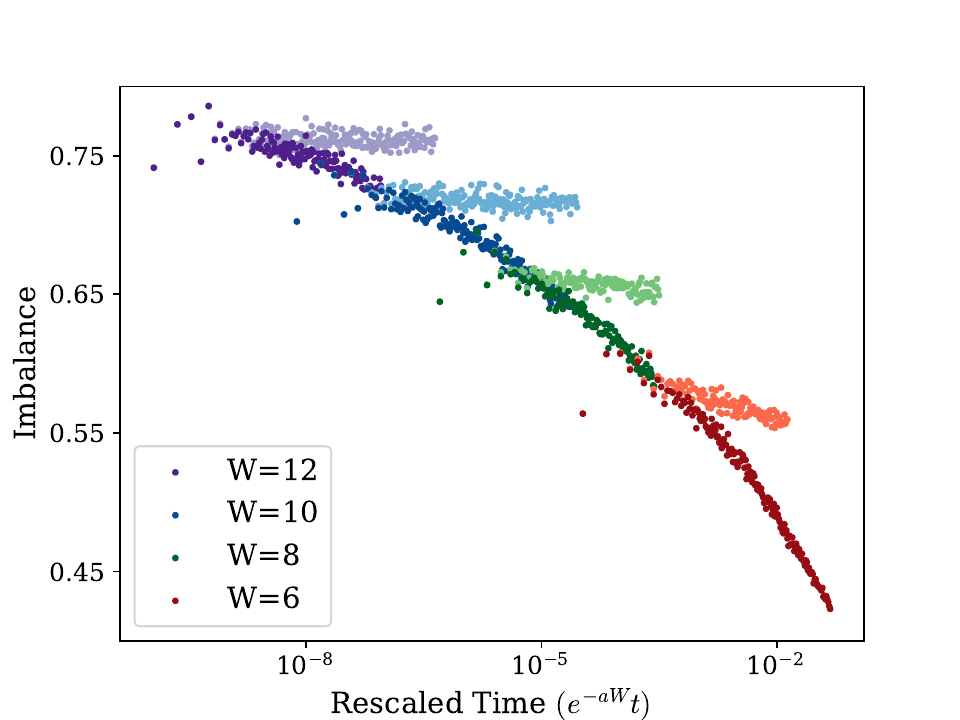}
\caption{\label{fig:all} 
Imbalance calculated in the middle of the strongly disordered XXZ chain, with (bold color) and without (light color) an added thermal inclusion, for different disorder strengths: from top to bottom is $W=12,10,8,6$. The bottom plot shows these results with the time axis rescaled by $t\rightarrow t e^{-aW}$ with $a=2.1$. All results are averaged over 400 disorder realizations.}
\end{figure}

The imbalance for both the biased and unbiased sampling, and various disorder strengths $W$ is shown in Fig. \ref{fig:all}. In absence of a rare region the imbalance decays very slowly at disorder $W=6-8$, but decay becomes indistinguishable from the sampling noise at larger disorder $W=10-12$. These results are in complete agreement with recent results in Ref.~\cite{Challenges,Doggen2018} and consistent with recent studies of the imbalance in similar finite systems \cite{Luitz2016ExtendedRegime,Chanda2020}. However, in the presence of the rare region the imbalance drops consistently up to our longest accessible time, after remaining steady for the first $\approx100/J$. With increasing disorder the the effect of the thermal inclusion sets in at later times, indicating progressively slower but still steady thermalization of the system. Furthermore, we find that in the presence of the thermal inclusion the imbalance seems to have some universal simple scaling behavior, i.e. simply expression time in scaled units $t'= te^{-aW}$ where $a$ is some constant of order one, results in a nice collapse of the data. We found the collapse to occur at $a\approx2.1$, which is shown in Fig. \ref{fig:all}. This collapse indicates that there are no remarkably different physics when the disorder is cranked up to $W=12$, rather the same physics operating at differing timescales which are set by the disorder strength. \par

Thus from these results, though still limited in system size and time, there is no sign of an MBL phase which is robust against Griffith's effects from rare thermal regions, even at large disorder. Instead, we see steady thermalization of the large disorder chain as it is increasingly entangled with the rare region. This result substantiates recent predictions from Refs.~\cite{MorningstarAvalanches,Bath-induceddelocalization} based on the coupling of an ideal Markovian bath. It's also worth noting that similar behavior has been observed in classical disordered systems without the need for explicitly sampling rare regions~\cite{sajna23}; see also Fig. 3.2 in Ref.~\cite{jon202}.   

\section{Brickwork}
\begin{figure}[t]
\includegraphics[width= 0.45\textwidth]{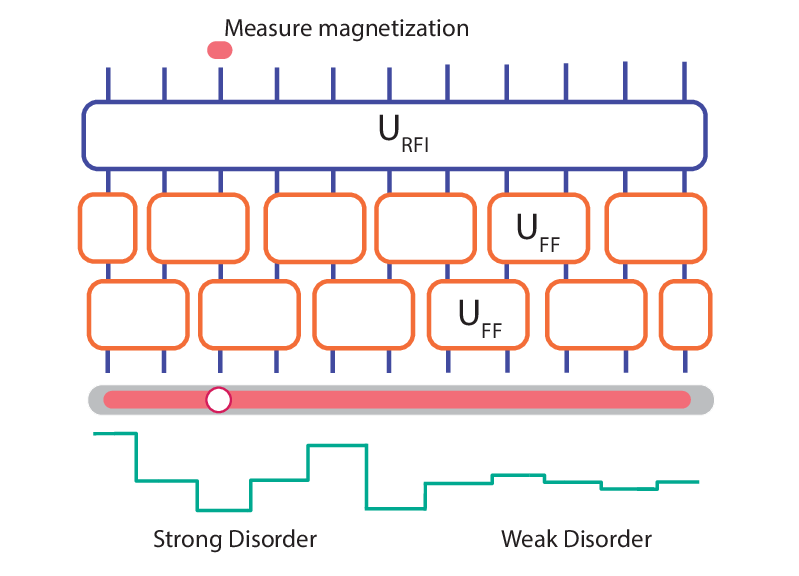}
\caption{\label{fig:diagram2} Diagram depicting the general setup, where we connect a region with strong disorder to a region with much weaker disorder. The system is initialized in a Haar random state, after which we project one of the spins in the strong disorder region to be up. We evolve the system in a brickwork circuit fashion, by layering flip-flop gates with random field Ising gates. We measure the residual magnetization on the probe spin.}
\end{figure}

\begin{figure}[t]
\includegraphics[width= 0.48\textwidth]{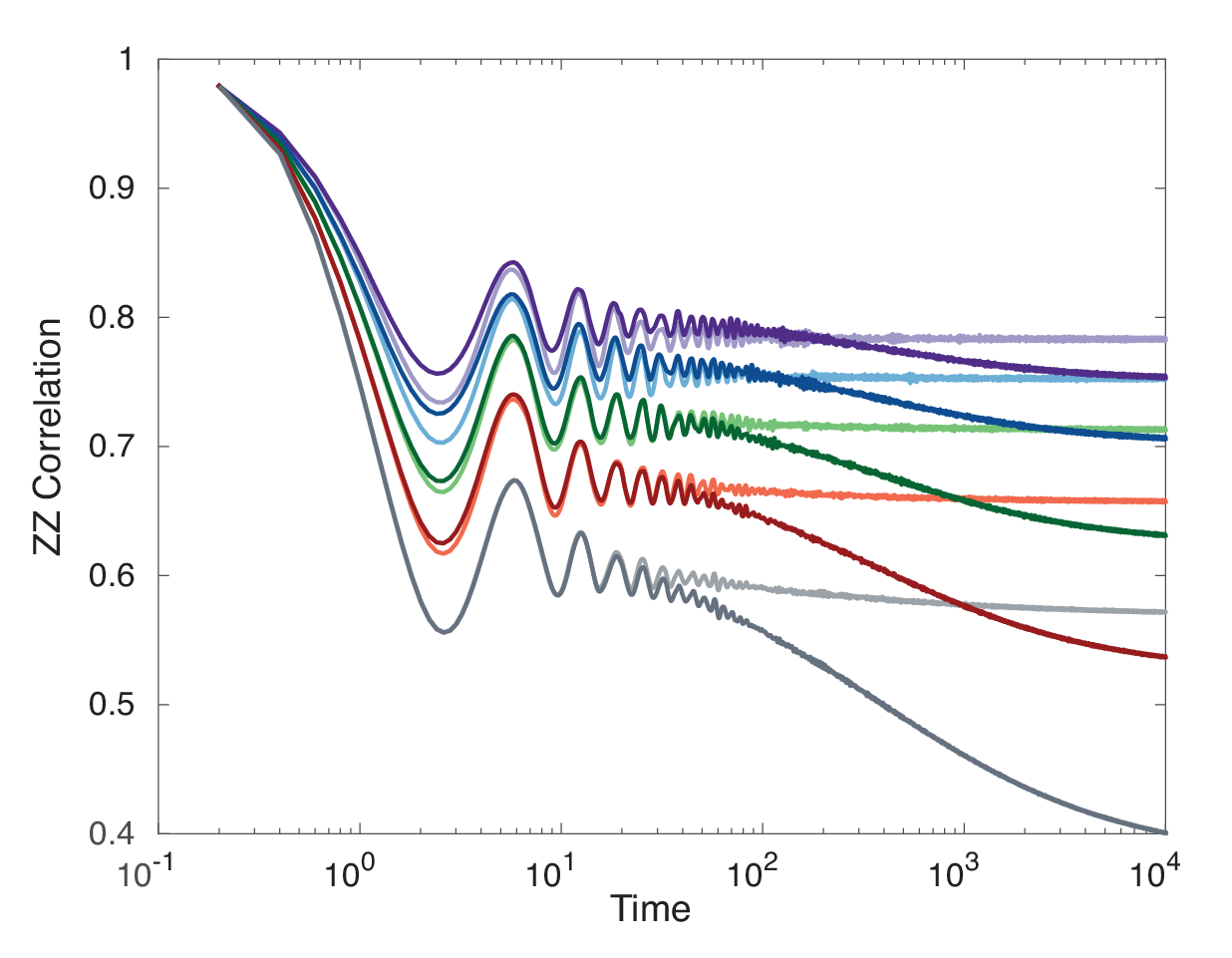}
\includegraphics[width= 0.48\textwidth]{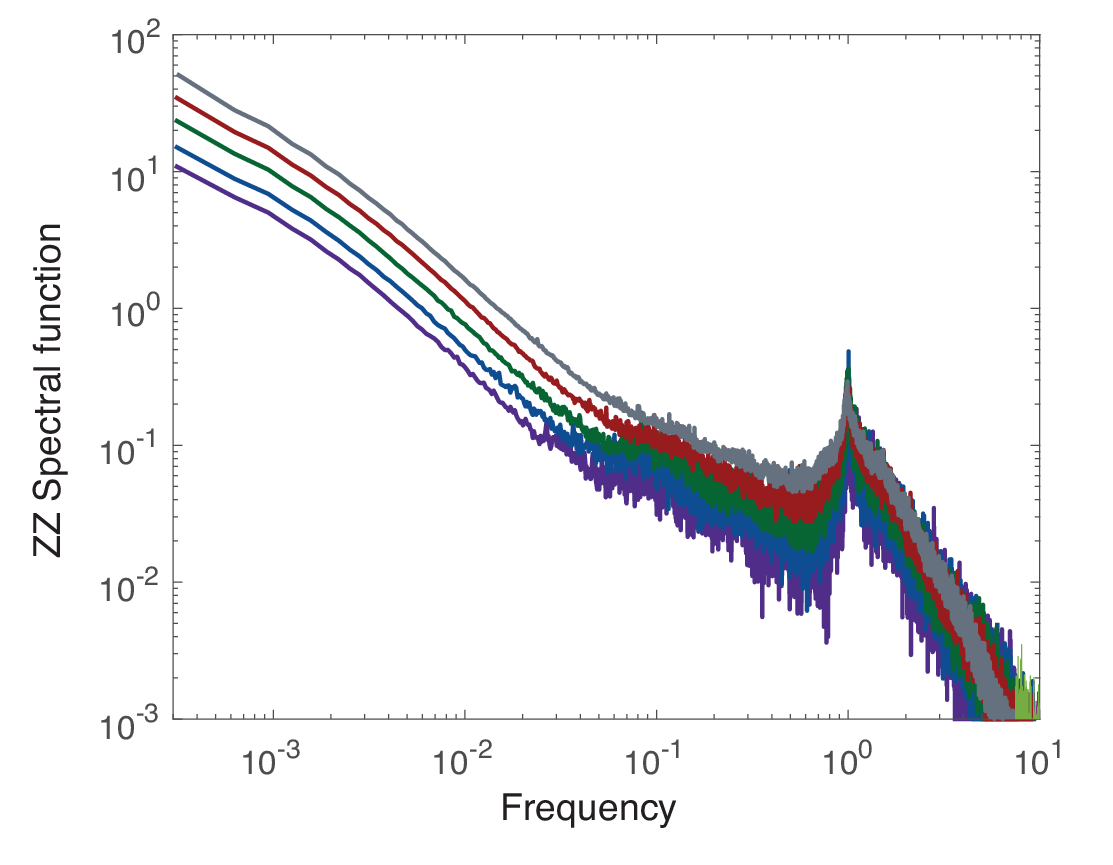}
\caption{\label{fig:spectrum} Two-time correlation function of a local spin in the bulk of a typical strongly disordered region. Panel A shows the correlation function for disorder ranging from $W=4,5,6,7,8$, with and without an embedded rare region. Panel B shows the spectral function, i.e. the Fourier transform of the correlation function, in the presence of a rare region. All results are averaged over 500 disorder realizations.}
\end{figure}

To make sure that the observed effect is not a consequence of the particular block construction nor of the fact that the state is being compressed into an MPS, we present a second method that keeps track of the exact state and evolves the system in a brickwork approximation of the Heisenberg model, as shown in Fig.~\ref{fig:diagram2}. Specifically, the evolution is split into a random field Ising part and a set of flip-flops: 
\begin{eqnarray}\label{bricks}
&&H_{\rm RFI}=\sum_{i=1}^{N-1} S^Z_{i}S^Z_{i+1} + W\sum_{i=1}^{N}h_iS^z_i  \nonumber \\
&&H_{\rm FF}=\sum_{i=1}^{N-1} \left(S^X_{i}S^X_{i+1}+S^Y_{i}S^Y_{i+1}\right)
\end{eqnarray}
The flip-flop are applied on all the even and all the odd bonds consecutively and all unitaries are constructed using a timestep $\tau=0.1$, see Fig.~\ref{fig:diagram2}. Since the exact state is kept in the time-evolution there is no need to restrict to initial states that are weakly entangled. Consequently, we can efficiently extract the infinite temperature spectral function of local observables by approximating the ensemble average by an expectation value in a Haar random state. Consider the ZZ-correlation function of the $n$th spin:
\begin{equation}
    S(t)=\frac{{\rm Tr} \left[\sigma^Z_{n}(t)\sigma^Z_{n}(0) \right]}{{\rm Tr}  \mathbb{I}},
\end{equation}
Using the fact that all Pauli's are traceless and projectors are idempotent, this can be rewritten as:
\begin{equation}
    S(t)=\frac{{\rm Tr} \left[ P_{\uparrow} \sigma^Z_{n}(t) P_{\uparrow} \right]}{{\rm Tr}  \mathbb{I}}=\mathbb{E_\psi} \left<\psi_{\uparrow}(t)\right| \sigma^z \left| \psi_\uparrow(t) \right> 
    \label{eq:haar}
\end{equation}
where $P_{\uparrow}=(1+\sigma^z_n)/2$ projects the $n$th spin to the up state, $\left| \psi_\uparrow(t) \right> =U_t P_{\uparrow} \left| \psi \right>$, and the average should be understood as averaging over Haar random states $ \left| \psi \right>$ (see for example also expressions S1-S6 in Ref. \cite{Richter2021RandomCircuits} for further explanation). Because the variance of the estimator in expression~\eqref{eq:haar} is suppressed by the Hilbert space dimension, one needs very few samples to obtain a good estimate of $S(t)$. In the results that follow we restrict ourselves to a single sample, since we average over 500 disorder realizations anyway.
The results are summarized in Fig.~\ref{fig:spectrum}, which shows the correlation function $S(t)$ for a spin in the middle of the strong disorder region. Results are presented for disorder ranging from $W=4-8$ for systems with and without biased sampling. In absence of rare regions there is some some slow decay for the weaker disorder values, while it becomes hard to distinguish decay from the noise over the accessible time at large disorder. As before, we observe a pronounced decay in the correlation function when the rare region is present. Multiple timescales seem to be involved in the problem from a damped oscillation at frequency $J$, coming from the local flip-flops to a slow decay at long times. This makes it hard to come up with a nice scaling function. Nonetheless, it seems the spectral function exhibits some universal behavior in this regime, see lower panel of Fig.~\ref{fig:spectrum}. 

\section{\label{sec:level5}Conclusions}
We present evidence that localization in small systems is not robust to thermal inclusions at much larger disorder than what is estimated from numerics using only unbiased sampling. The existence of a rare region begets the start of a quantum avalanche, in which the surrounding spins get entangled, growing the rare region and setting off a steady thermalization of the entire system. This study illustrates the importance of including rare regions in studying the stability of the MBL phase in the thermodynamic limit, and it reflects recent results from a cold-atom experiment on smaller system sizes~\cite{Leonard2023}. Over the studied range of disorder, we find no qualitative change in the behavior of the system. In contrast, it appears that all systems behave the same as long as one accounts for an exponential slowdown of the thermalization with disorder. As usual, this study is limited in system size and simulation times due to the high complexity of the many body problem.

\emph{Acknowledgments.}
The Flatiron Institute is a division of the Simons Foundation. We acknowledge support from AFOSR: Grant FA9550-21-1-0236.

\appendix

\section{Appendix A: TEBD construction}
\begin{figure}[t]
\includegraphics[width=0.48\textwidth]{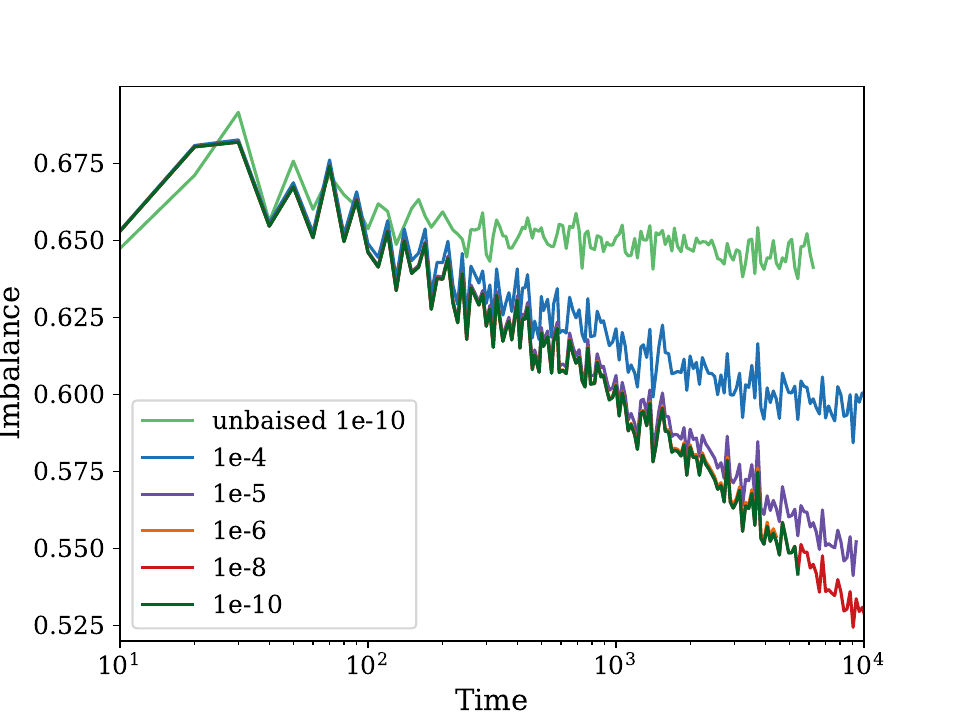}
\caption{\label{fig:Cutoff}
Imbalance in the disordered XXZ chain for $W=8$, with the rare-region thermal inclusion attached except for where specified in the top curve, for different ITensor cutoff values. Results are averaged over 200 disorder realizations.}
\end{figure}

The results in the Block Evolution section were obtained by a time-evolving block decimation algorithm (TEBD) for matrix product states (MPS) as implemented in the Julia ITensor library \cite{itensor}. Instead of evolving with a cascade of two-site matrix product operator's (MPO's) as is typical for TEBD, we construct one larger MPO for each chain, with a third two-site MPO connecting the two chains. This approach allows to use much larger time steps of $dt=10$ because each block, being $L_{MBL} = 12$ and $L_{RR}=10$ sites respectively, is small enough to be evolved quasi-exactly with precision controlled by ITensor ``cutoff'' parameter. We find that our results converge with a cutoff greater than $10^{-6}$, as shown in Fig. \ref{fig:Cutoff}, and throughout the work we use a cutoff of $10^{-10}$. It should be noted that the bias for unconverged results is to larger imbalance values. The latter is expected when using TEBD to study MBL-type dynamics, and similar results were obtained in Ref.~\cite{Challenges}.\par 

\nocite{*}

\bibliography{apssamp}

\end{document}